\documentclass[aps,amsmath,amssymb,prl,twocolumn,superscriptaddress,showpacs]{revtex4}
\usepackage{bm}
\usepackage{epsfig}
\usepackage[usenames]{color}
\usepackage{graphicx}

\renewcommand{\paragraph}[1]{\textit{#1.---} }
\newcommand{\paper}{paper }

\begin{document}

\title{Nanogranular Thermoelectrics}

\author{Andreas~Glatz}
\affiliation{Materials Science Division, Argonne National Laboratory, Argonne, Illinois 60439, USA}

\author{I.~S.~Beloborodov}
\affiliation{Department of Physics and Astronomy, California State University Northridge, Northridge, CA 91330, USA}

\date{\today}

\begin{abstract}
We investigate thermopower and thermoelectric coefficient of nano-granular materials at large tunneling conductance between the grains, $g_{T}\gg 1$.
We show that at intermediate temperatures, $T\geq g_{T}\delta$, where $\delta$ is the mean energy level spacing for a single grain, electron-electron interaction leads to an increase of the thermopower with decreasing grain size. We discuss our results in the light of new types of thermoelectric materials and present the behavior of the figure of merit depending on system parameters.
\end{abstract}

\maketitle

The search for more efficient thermoelectric materials has had little success during the last several decades since bulk materials are limited in their performance by the Wiedemann-Franz law that connects electric to thermal conductivity in such a way as to defeat all attempts at improving the dimensionless figure of merit $ZT = S^2\sigma T/\kappa$, where $S$ is the thermopower or Seebeck coefficient, $\sigma$ is the electric conductivity, and $\kappa$ is the thermal conductivity~\cite{Rowe,Abrikosov,Mahanbook,Mahan, Bell}.
To be competitive compared with conventional refrigerators, one must develop thermoelectric materials with $ZT > 2$. Although it is possible in principle to develop homogeneous materials with $ZT > 2$, there are no candidate materials on the horizon.  Thus, one needs to search for {\it inhomogeneous/granular} thermoelectric materials in which  one can directly control the system parameters.

Most theoretical progress was archived by numerical solution of phenomenological models~\cite{Linke,Broido}. However, no analytical results obtained from a microscopic model for coupled nanodot/grain systems is available till now. Thus, the fundamental question that remains open is how thermoelectric coefficient and thermopower behave in nanogranular thermoelectric materials.  Here, we make the first step towards answering this question for granular metals at intermediate temperatures by generalizing our approach~\cite{Beloborodov07} recently developed for the description of electric~\cite{Beloborodov03} and heat transport~\cite{Beloborodov05}.  In particular, we will answer the question to what extend quantum and confinement effects in nanostructures are important in changing $ZT$.

In this \paper we investigate the thermopower $S$, thermoelectric coefficient $\eta$, and the figure of merit $ZT$ of granular samples focusing on the case of large tunneling conductance between the grains, $g_{T}\gg 1$.
Without Coulomb interaction the granular system would be a good metal in this limit and our task is to include charging effects in the theory.

\paragraph{Main results} The main results of our work are as follows: (i)  We derive the expression for the thermoelectric coefficient $\eta$ of granular metals that includes corrections due to Coulomb interaction at temperatures $T> g_T \delta$, where $\delta$ is the mean level spacing of a single grain
\begin{equation}\label{eq.eta}
\eta =\eta ^{(0)}\left( 1-\frac{1}{4 g_{T} d} \, \ln \frac{g_T E_c}{T} \right).
\end{equation}
Here $ \eta^{(0)} = - (\pi^2/3)e g_T a^{2-d} (T/\varepsilon _{F})$ is the thermoelectric coefficient
of granular materials in the absence of electron-electron interaction with $e$ being the electron charge, $a$ the size of a single grain, $d= 2,3$ the dimensionality of a sample, $\varepsilon_F$ being the Fermi energy, and $E_c=e^2/a$ is the charging energy.
\begin{figure}[b]
\includegraphics[width=0.8\columnwidth]{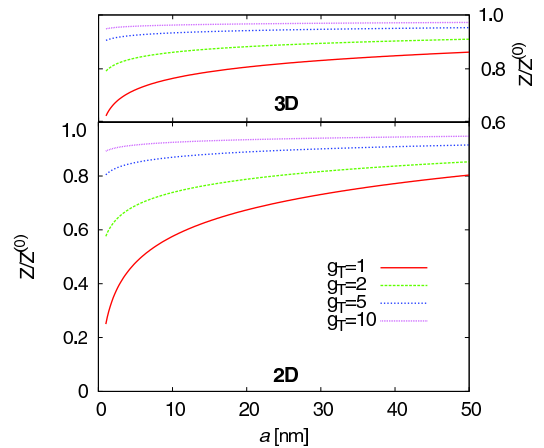}
\caption{Plots of the dimensionless figure of merit $Z/Z^{(0)}$  vs. grain size $a$ (in nm)- for different values of dimensionless tunneling conductance $g_T$ (see legend): the upper panel is for 3D and the lower for 2D. All curves are plotted for $T=100$K. At this temperature, the dimensionless bare figure of merit for granular metals is $Z^{(0)}T\approx 10^{-4}$.}
\label{fig.plots}
\end{figure}

(ii) We obtain the expression for thermopower $S$ of granular metals
\begin{equation}\label{eq.S}
S = S^{(0)}\left(1 - \frac{\pi -2}{4 \pi g_T d} \ln \frac{g_T E_c}{T} \right),
\end{equation}
where $S^{(0)} = - (\pi^2/6) (1/e) (T/\varepsilon_F)$ is the thermopower of granular metals in the absence of Coulomb interaction.

(iii) Finally, we find the figure of merit to be:
\begin{equation}\label{eq.ZT}
\frac{Z}{Z^{(0)}} = 1 - \frac{\pi - 2}{2 \pi g_T d}  \ln \frac{g_T E_c}{T} -
\frac{1}{2\pi^2 g_T}\left\{
\begin{array}{lr}
 3 \gamma,
\hspace{0.8cm} d=3  \\
\ln \frac{g_{T} E_c }{T}, \hspace{0.1cm} d=2
\end{array}
\right.,
\end{equation}
where $Z^{(0)} T = (\pi^2/12) (T/\varepsilon_F)^2$ is the bare figure of merit of granular materials and $\gamma \approx 0.355$ is a numerical coefficient.
In Fig.~\ref{fig.plots} we plot $Z/Z^{(0)}$ as a function of the grain size $a$ for different tunneling conductances $g_T$ at fixed temperature $100$K.  We find, that the influence of granularity is most effective for small grain sizes and the presence of Coulomb interaction decreases the figure of merit.

At this point we remark that all results are obtained in the absence of phonons which become relevant only at higher temperatures. At the end of this \paper we will briefly discuss their influence.

Our main results, Eqs.~(\ref{eq.eta}) - (\ref{eq.ZT}), are valid at intermediate temperatures, $T > g_T \delta$. At these temperatures the electronic motion is coherent
within the grains, but coherence does not extend to scales
larger than the size $a$ of a single grain~\cite{Beloborodov07}. Under these
conditions, the electric conductivity $\sigma$ and the electric thermal conductivity $\kappa$ are given by the expressions~\cite{Beloborodov03,Efetov02,Beloborodov05}
\begin{subequations}
\begin{eqnarray}
\frac{\sigma}{\sigma^{(0)}} &=& 1 - \ln(g_TE_c/T)/(2\pi d g_T),
\label{eq.sigma}\\
\frac{\kappa}{\kappa^{(0)}} &=&  1 - \frac{\ln [g_T E_c/T]}{2\pi d g_T} + \frac{1}{2\pi^2 g_T}\left\{
\begin{array}{lr}
 3\gamma,
\hspace{0.8cm} d=3 &  \\
\ln \frac{g_{T} E_c }{T}, \hspace{0.1cm} d=2
\end{array}
\right. . \label{eq.kappa}
\end{eqnarray}
\end{subequations}
where $\sigma^{(0)} = 2 e^{2}g_{T}a^{2-d}$ and $\kappa^{(0)} = L_0 \sigma^{(0)} T$
are the electric (including spin) and thermal conductivities of granular metals in the absence of Coulomb interaction with $L_0 = \pi^2/3 e^2$ being the Lorentz number. We mention that at temperature $T > g_T \delta$ the correction to the thermoelectric coefficient, Eq.~(\ref{eq.eta}), has a $T \ln T$
dependence in both $d=2,3$ dimensions which is similar to the result for the electric conductivity, Eq.~(\ref{eq.sigma}), having a $\ln T$ dependence in all dimensions as well.

\paragraph{Model} Next we introduce our model and describe the derivation of
Eqs.~(\ref{eq.eta})-(\ref{eq.ZT}):  We consider a $d-$dimensional array of
metallic grains with Coulomb interaction between electrons.
The motion of electrons inside the grains is diffusive and they
can tunnel from grain to grain. We assume that the sample would be a good metal in the absence of Coulomb interaction.  However, we also assume that $g_{T}$ is still smaller than the grain conductance $g_{0}$,
meaning that the granular structure is pronounced and the
resistivity is controlled by tunneling between grains.

Each grain is characterized by two energy scales: (i) the mean
energy level spacing $\delta$, and (ii)
the charging energy $E_c = e^2/a$ (for a typical grain size of $a \approx
10$nm $E_c$ is of the order of $2000$K) and we assume that the condition
$\delta \ll E_c$ is fulfilled.

The system of coupled metallic grains is described by the Hamiltonian $\hat{H} =
\sum _i \hat{H}_i$, where the sum is taken over all grains in the system and
\begin{equation}
\label{Hamiltonian}
\hat{H}_i =  \sum_{k}\xi_k \hat{a}^{\dagger}_{i,k} \hat{a}_{i,k} + \sum_{j \ne i}\,\frac{e^{2}\hat{n}_{i}\hat{n}_{j}}{2C_{ij}}\,
 + \sum_{j,p,q} (t_{ij}^{pq} \hat{a}^{\dagger}_{i,p} \hat{a}_{j,q} +\text{c.c.}) .
\end{equation}
The first term in the right hand side (r.~h.~s.)
of Eq.~(\ref{Hamiltonian}) describes the $i$-th isolated disordered grain, $\hat{a}^{\dagger}_{i,k} (\hat{a}_{i,k})$ are the creation
(annihilation) operators for an electron in the state $k$ and $\xi_k = k^2/2m - \mu$
with $\mu$ being the chemical potential. The second term describes the charging
energy, $C_{ij}$ is the capacitance matrix and $\hat{n}_{i} = \sum_k \hat{a}^{\dagger}_{i,k} \hat{a}_{i,k}$ is the number operator for
electrons in the $i$-th grain. The last term is the tunnel Hamiltonian
where $t_{ij}$ are the tunnel matrix elements between grains $i$ and $j$.

\begin{figure}[b]
\includegraphics[width=0.6\columnwidth]{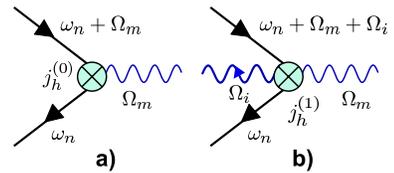}
\caption{Vertices corresponding to the thermal current operator, Eqs.~(\ref{thermalvertex}):
vertex (a) corresponds to $\widehat{\jmath }_{ij}^{(h,0)}$ and (b)
to $\widehat{\jmath }_{ij}^{(h,1)}$. The solid lines denote the propagator of electrons, the thick wavy line describes Coulomb interaction, the tunneling vertices are described by the circles, $\omega_n = \pi T(2n+1)$ and $\Omega_m =2\pi m T$ are Fermionic and Bosonic Matsubara frequencies respectively ($n,m\in\mathbb{Z}$).}
\label{fig.jver}
\end{figure}

\paragraph{Derivation of kinetic coefficients} The kinetic coefficients: electric conductivity $\sigma$, the thermoelectric coefficient $\eta$, and the thermal conductivity $\kappa$ are related to the Matsubara response functions $L^{(\alpha\beta)}$ with $\alpha,\beta\in\{e,h\}$~\cite{Mahan,Mahanbook,Jonson}
\begin{eqnarray}
{\bf j}^{(e)}& =& - (L^{(ee)}/(e^2 T))\, \nabla (e V) - (L^{(eh)}/(e T^2))\,  \nabla T, \nonumber \\
{\bf j}^{(h)} &=& - (L^{(eh)}/(e T))\, \nabla (e V) - (L^{(hh)}/T^2) \, \nabla T.
\label{responce}
\end{eqnarray}
Here ${\bf j}^{(e)}$ (${\bf j}^{(h)}$) is the electric (thermal) current and $V$ the electrostatic potential.
From Eq.~(\ref{responce}) one finds that $\sigma = L^{(ee)}/T $, $\eta = L^{(eh)}/T^2$, and $S = -\Delta V / \Delta T = L^{(eh)}/(T L^{(ee)})$, where the response functions are given by Kubo formulas $L^{(\alpha\beta)} =  -\left.\frac{\imath T \, \partial }{a^{d} \partial \Omega }\right|_{\Omega \rightarrow 0}  \left[ \int\limits_0^{1/T}d\tau \, e^{\imath\Omega_m \tau} \langle T_{\tau}{\bf j}^{(\alpha)}(\tau) {\bf j}^{(\beta)}(0)\rangle \right]_{\Omega_m \rightarrow - \imath\Omega + \delta}$, with $T_{\tau}$ being the time ordering operator for the currents with respect to the imaginary time $\tau$. Thus, to calculate the thermoelectric coefficient
$\eta$ and thermopower $S$ of the granular metals one has to know
the explicit form of the electric ${\bf j}^{(e)}$ and thermal ${\bf j}^{(h)}$ currents.

The electric current ${\bf j}_i^{(e)}$ through grain $i$ is defined as
${\bf j}^{(e)}_i = \sum_j \hat{\jmath }_{ij}^{(e)} = e\, d\hat{n}_{i}/dt = \imath e[\hat{n}_{i}, \hat{H]}$. Straightforward calculations lead to $\hat{\jmath }_{ij}^{(e)}=\imath e\underset{k,q}{\sum }
( t_{ij}^{kq} \hat{a}_{i,k}^{\dagger } \hat{a}_{j,q}-t_{ji}^{qk} \hat{a}_{j,q}^{\dagger } \hat{a}_{i,k})$.

For granular metals the thermal current operator ${\bf j}_i^{(h)} = \sum_j \hat{\jmath }_{ij}^{(h)}$ can be obtained as follows. The energy content of each grain changes as a function
of time, such that $d\hat{H}_i/dt = i [\hat{H}_i,\hat{H}]$. Energy conservation requires that this energy flow to the other grains in the system, $d\hat{H}_{i}/dt \equiv \sum_j \hat{\jmath }_{ij}^{(h)}$. Calculating the
commutator $[\hat{H}_i,\hat{H}]$, we obtain $\hat{\jmath }_{ij}^{(h)} = \hat{\jmath }_{ij}^{(h,0)} + \hat{\jmath }_{ij}^{(h,1)}$, where
\begin{subequations}\label{thermalvertex}
\begin{eqnarray}
\hat{\jmath }_{ij}^{(h,0)}&=&\imath \underset{k,q}{\sum }
\frac{\xi _{k}+\xi _{q}}{2}\left[ t_{ij}^{kq} \hat{a}_{i,k}^{\dagger }
\hat{a}_{j,q}-t_{ji}^{qk} \hat{a}_{j,q}^{\dagger } \hat{a}_{i,k}
\right],\\
\hat{\jmath }_{ij}^{(h,1)}&=&-\frac{e}{4}\underset{m}{
\sum } \left[ \frac{\{ \hat{n}_{i};\hat{\jmath }
_{jm}^{(e)} \} _{+}}{C_{im}}-\frac{ \{ \hat{n}_{j};\hat{\jmath }
_{im}^{(e)} \} _{+}}{C_{jm}} \right],
\end{eqnarray}
\end{subequations}
where $\{ \hat{A};\hat{B} \} _{+}$ stands for the anti-commutator.
The contribution $\widehat{\jmath }_{ij}^{(h,0)}$ is the heat current in the absence of electron-electron interaction, while the second term $\hat{\jmath }_{ij}^{(h,1)}$ appears due to Coulomb interaction. Equations~(\ref{thermalvertex}) imply that the thermal current operator must be associated with two different vertices in diagram representation, Fig.~\ref{fig.jver}.

For large tunneling conductance, the Matsubara thermal
current - electric current correlator can be analyzed perturbatively in
$1/g_T$, using the diagrammatic technique discussed in
Ref.~\onlinecite{Beloborodov07} that we briefly outline
below. The self-energy of the averaged single electron Green
function has two contributions: The first contribution corresponds
to scattering by impurities inside a single grain while the second
is due to processes of scattering between the grains. The former
results only in small renormalization of the mean free time which depends in general on the electron energy $\omega$ as $\tau _{\omega }^{-1} = \tau _{0}^{-1}\left[ 1+(d/2-1)\omega /\varepsilon _{F}\right]$ which is a result of the renormalization of the density of states at the Fermi surface.

The diffusion motion inside a single grain is given by the
diffusion propagator $ {\cal D}_0^{-1} = \tau_{\omega}|\Omega_i|$,
where $\Omega_i$ is the bosonic Matsubara
frequency for the (internal) Coulomb interaction. The coordinate dependence in ${\cal D}_0$ is neglected since
in the regime under consideration all characteristic energies are
smaller than the Thouless energy. The complete diffusion propagator
is given by ladder diagrams resulting in the following expression ${\cal D}^{-1}(\Omega_i, {\bf q}) = \tau_{\omega}(|\Omega_i|+
\epsilon_{\bf q} \delta )$,
where $\epsilon _{\bf q}=2 g_{T}\sum_{\mathbf{a}}(1-\cos
\mathbf{qa})$ with $\mathbf{a}$ being the lattice vectors.
The same ladder diagrams describe the renormalized interaction
vertex. The interaction vertex is used to obtain the polarization operator,
that defines the effective dynamically screened Coulomb interaction for granular metals,
$V(\Omega_i, {\bf q})=2\left[ E^{-1}_c({\bf q}) + 4\epsilon_{
\bf q}/(|\Omega_i| +  \epsilon_{\bf q}\delta)  \right]^{-1}$,
where the charging energy in $d=2,3$ is given by
$E_c({\bf q})=\frac{e^{2}}{2C({\bf q})}=\frac{2^d\pi e^{2} q}{4(aq)^{d}}$.

\begin{figure}[t]
\includegraphics[width=\columnwidth]{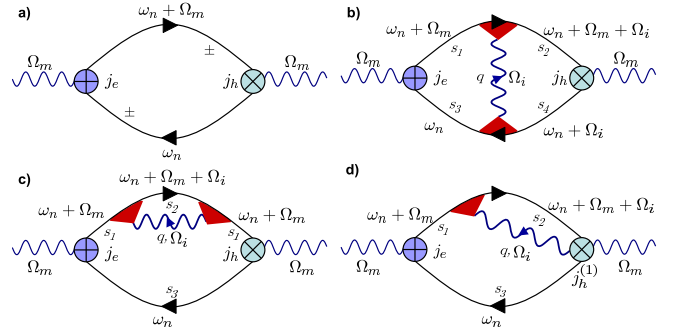}
\caption{Diagrams describing the thermoelectric
coefficient of granular metals at temperatures $T > g_T\delta$:
diagram (a) corresponds to $\protect\eta_{0}$
in Eq.~(\ref{eq.eta}). Diagrams (b)-(d) describe first order corrections
to the thermoelectric coefficient of granular metals due to electron-electron
interaction. The solid lines denote the propagator of electrons,
the wavy lines describe effective screened electron-electron
propagator, and the (red) triangles describe the elastic
interaction of electrons with impurities. The tunneling vertices are described by the circles.
The sum of the diagrams (b)-(d) results in the thermoelectric coefficient
correction $\eta^{(1)}$ given in Eq.~(\ref{eta1}).}
\label{fig.dia}
\end{figure}
However, there is an important difference between calculations of thermoelectric coefficient $\eta$ and thermopower $S$ and the calculations of the electric $\sigma$ and thermal $\kappa$ conductivities in Eqs.~(\ref{eq.sigma}, \ref{eq.kappa}). Indeed, to calculate $\sigma$ and $\kappa$ it was sufficient to approximate the tunneling matrix element $t_{pq}$ as a constant $t$ which is evaluated at the Fermi surface and neglect variations of $t_{pq}$ with energy which occur on the scale $T/\varepsilon_F$. However, this approximation is insufficient for calculations of thermoelectric coefficient $\eta$ and thermopower $S$ since the dominant contribution to these quantities vanishes due to particle-hole symmetry such that that both quantities are proportional to the small parameter $T/\varepsilon_F$. Since it is necessary to take into account terms of order of $T/\varepsilon_F$ in order to obtain a nonzero result for $\eta$ and $S$ the corresponding expansions must be carried out to this order for all quantities which depend on energy: the density of states, the relaxation time, and the tunneling matrix element. For the latter we obtain the expression: $t^{2}(\xi _{1},\xi _{2}) = t_{0}^{2} \left( 1+\frac{\xi _{1}+\xi _{2}}{ \varepsilon _{F}}\right)$~\footnote{The energy dependent expression for the tunneling matrix element includes the effect of multi-channel tunneling.}.

In the absence of the electron-electron interaction the thermoelectric coefficient is
represented by diagram (a) in Fig.~1. Straightforward calculations of this diagram lead to the result for $\eta^{(0)}$ given below Eq.~(\ref{eq.eta}).

First order interaction corrections to the thermoelectric coefficient are only generated by diagrams (b) and (c) in Fig.~\ref{fig.dia}, resulting after summation over Fermionic and Bosonic frequencies and analytical continuation in
\begin{equation}
\label{eta1}
\eta ^{(1)} = -\frac{\eta ^{(0)}}{2\pi g_{T}}\left( \frac{a}{2\pi }\right) ^{d}\int d^d {\bf q} \ln \left[\frac{2E_c({\bf q})\epsilon_{{\bf q}}}{T}\right],
\end{equation}
where the ${\bf q}$-integration goes over the $d$-dimensional sphere with radius $\pi/a$. Diagram (d) in Fig.~\ref{fig.dia} gives only contributions to the thermoelectric coefficient of order $(T/\varepsilon_F)^2$ and higher.
Integrating over ${\bf q}$ in Eq.~(\ref{eta1}) we obtain the following expressions:
\begin{equation}
\eta _{2d}^{(1)} = -\frac{\eta ^{(0)}}{8 g_{T}}
\ln \frac{E_c g_{T}}{T}, \hspace{0.5cm}
\eta _{3d}^{(1)} = -\frac{\eta ^{(0)}}{12 g_{T}}\ln \frac{E_c g_{T}}{T},
\end{equation}
which lead to Eq.~(\ref{eq.eta}).

\paragraph{Discussions} In the presence of interaction effects and not very low temperatures $T > g_T \delta$, granular metals behave differently from homogeneous disorder metals. However, in the absence of interactions the result for $\eta^{(0)}$ below Eq.~(\ref{eq.eta}) coincides with the thermoelectric coefficient of homogeneous disordered metals, $\eta^{(0)}_{hom} = -(2/9)e p_F (\tau_0 T)$, with $p_F$ being the Fermi momentum. One can expect that at low  temperatures, $T < g_T \delta$, even in the presence of Coulomb interaction the behavior of thermoelectric coefficient and thermopower of granular metals is similar to the behavior of $\eta_{hom}$ and $S_{hom}$, however this temperature range is beyond the scope of the present \paper.
Our results for thermopower (\ref{eq.S}) and figure of merit (\ref{eq.ZT}) show that the influence of Coulomb interaction is most effective for small grains. $S^2$ decreases with the grain size which is a result of the delicate competition of the corrections of thermoelectric coefficient (\ref{eq.eta}) and the electric conductivity (\ref{eq.sigma}). In particular, if the numerical prefactor of the correction to $\eta$ would be slightly smaller, the sign of the correction to $S$ would change.

Above we only considered the electron contribution to figure of merit. At higher temperatures $T>T^*$, where $T^* \sim \sqrt{g_T c_{ph}^2/l_{ph}} a$ is a characteristic temperature with  $l_{ph}$ and $c_{ph}$ being the phonon scattering length and phonon velocity respectively~\cite{Beloborodov05}, phonons will provide an independent, additional contribution to thermal transport, $\kappa_{ph} = T^3 l_{ph}/c_{ph}^2$. However, the phonon contribution can be neglected for temperatures $g_T \delta < T < T^*$. A detailed study of the influence of phonons at high temperatures, including room temperature, will be subject of a forthcoming work.

So far, we ignored the fact that electron-electron interactions also renormalize the chemical
potential $\mu$. In general, this renormalization may affect the kinetic coefficients: the thermal current vertex, Fig~\ref{fig.jver}, as well as the electron Green functions depend on $\mu$. In particular one needs to replace $\nabla (e V) \rightarrow \nabla (e V + \mu)$ in Eqs.~(\ref{responce}). To first order in the interactions, the renormalization of $\mu$ only leads to corrections to diagram (a) in Fig.~\ref{fig.dia}. As it can be easily shown, for this diagram the renormalization of the heat and electric current vertices is exactly canceled by the renormalization of the two electron propagators. Therefore, the renormalization of the chemical potential does not affect our results in the leading order.

Finally, we remark that the bare figure of merit $Z^{(0)}T$ for granular metals at $g_T > 1$ and $100$K is of the order of only $10^{-4}$. Therefore these materials are not suitable for {\it solid-state refrigerators}, but should be replaced by granular semiconductors with $g_T < 1$. Therefore we conclude this paragraph discussing the dimensionless figure of merit $ZT$ of granular materials at weak coupling between the grains, $g_T \ll 1$. In this regime the electronic contribution to thermal conductivity $\kappa_e$ of granular metals was recently investigated in Ref.~\onlinecite{Tripathi}, where it was
shown that $\kappa_e \sim g_T^2 T^3/E_c^2$.  In this regime the electric conductivity of granular metals obeys the law $\sigma \sim g_T \exp(-E_c/T)$~\cite{Efetov02,Beloborodov07}. However, an expression for the thermoelectric coefficient in this region is not available yet, but recently it has been proposed, based on experiment, that nanostructured thermoelectric materials in the low coupling region (AgPb$_m$SbT$e_{2+m}$, Bi$_2$Te$_3$/Sb$_2$Te$_3$, or CoSb$_3$)~\cite{Hsu,Poudel,Majumdar,Mi,Harman} can have higher figures of merit than their bulk counterparts.

In conclusion, we have investigated the thermoelectric coefficient and thermopower of granular nanomaterials in the limit of large tunneling conductance between the grains and temperatures $T > g_T\delta$. We have shown to what extend quantum and confinement effects in granular metals are important in changing $ZT$ depending on system parameters.

\paragraph{Acknowledgements} We thank Frank Hekking, Nick Kioussis, and Gang Lu for
useful discussions. A.~G. was supported by the
U.S. Department of Energy Office of Science under the Contract No. DE-AC02-06CH11357.

\end{document}